\documentclass[aps,pra, twocolumn,a4paper,showpacs]{revtex4}

\usepackage{graphicx}
\usepackage{revsymb}
\usepackage{dcolumn}
\usepackage{bm}

\begin{document}
\title{Reaching optimally oriented molecular states by laser kicks}

\author{D. Sugny}
\email{dominique.sugny@u-bourgogne.fr}
\author{A. Keller}
\email{arne.keller@ppm.u-psud.fr}
\author{O. Atabek}
\affiliation{Laboratoire de Photophysique Mol\'{e}culaire du CNRS,
Universit\'{e} Paris-Sud, B\^{a}t. 210 - Campus d'Orsay, 91405 Orsay Cedex,
France}
\author{D. Daems}
\email{ddaems@ulb.ac.be}
\affiliation{Center for Nonlinear Phenomena and Complex Systems, Universit\'{e}
Libre de Bruxelles, CP 231, 1050 Brussels, Belgium}
\author{C. M. Dion}
\affiliation{Department of Physics, Ume\aa \ University, SE-90187 Ume\aa, Sweeden}
\author{S. Gu\'{e}rin and H. R. Jauslin}
\affiliation{Laboratoire de Physique de l'Universit\'{e} de Bourgogne, UMR CNRS
5027, BP 47870, 21078 Dijon, France}
\begin{abstract}
We present a strategy for post-pulse molecular orientation aiming both at efficiency and maximal duration within a rotational period. We first identify the optimally oriented states which fulfill both requirements. We show that a sequence of half-cycle pulses of moderate intensity can be devised for reaching these target states.
\end{abstract}
\pacs{33.80.-b, 32.80.Lg, 42.50.Hz}
\maketitle

Molecular orientation plays a crucial role in a wide variety of applications extending from chemical reaction dynamics, to surface processing,
catalysis and nanoscale design \cite{brook,aoiz,seideman,stapel}. Static electric field \cite{sakai} and
strong non-resonant long laser pulses \cite{vrakking,guerin} have been shown to yield
adiabatic molecular orientation which disappears when the pulse is off. Noticeable orientation that persists after the end of the
pulse (and even under thermal conditions) is of special importance for
experiments requiring field-free transient orientation. It has recently been
shown that very short pulses combining a frequency $\omega$ and its second
harmonic $2\omega$ excite a mixture of even and odd rotational levels and have
the ability to produce such post-pulse orientation \cite{dion1}. But even more
decisive has been the suggestion to use half-cycle pulses (HCPs), that through
their highly asymmetrical shape induce a very sudden momentum transfer to the
molecule which orients under such a kick after the field is off \cite{dion2,malchholm}. 
Both the ($\omega+2\omega$) and the kick mechanisms have received a confirmation
from optimal control schemes \cite{dion3}. The caveat is that the post-pulse orientation is maintained for only short times. Recently, the use of a train of kicks to increase the
efficiency of the orientation has been suggested in optimal
control strategies \cite{dion3} and applied to molecular alignment \cite{Leibscher} and orientation of a 2D rotor \cite{averbukh}. However, due to the strength of the kicks used, only the efficiency of the process has been optimized, its duration decreasing strongly. In the present letter, we propose a control strategy using specially designed series of kicks delivered by short HCPs, that allows to significantly enhance the duration of the orientation, maintaining a high efficiency. Our construction is first based on the identification of target states which fulfill the previous requirement. These states are characterized by the fact that they only involve a limited number of the lowest lying rotational levels and that they maximize the orientation efficiency within the corresponding restricted rotational spaces. At a second stage, we show that these selected states can be reached by a train of kicks, acting at appropriately chosen times. The choice of the strength of the pulses (taken equal for simplicity), together with the total number of kicks allow to approach these target states with good accuracy. 
 
The time evolution of the molecule (described in a 3D rigid rotor approximation) interacting
with a linearly polarized field is governed by the time-dependent
Schr\"{o}dinger equation (in atomic units) 
\begin{equation} \label{eqn1}
i\frac{\partial}{\partial
t}\psi(\theta,\phi,t)=[BL^2-\mu_0F(t)\cos\theta]\psi(\theta,\phi,t) ,
\end{equation}
where $L$ is the angular momentum operator, $B$ the rotational constant, $\mu_0$
the permanent dipole moment and $F(t)$ the field amplitude. $\theta$ denotes 
the polar angle between the molecular axis and the polarization direction of the
applied field. The motion related to the azimuthal angle $\phi$ can be separated due to
cylindrical symmetry. From now on, we assume a sudden approximation due to 
the short durations $\tau$ of the HCPs, as compared to the molecular rotational
period $T_{{\rm rot}}=\pi/B$. For relatively low $l$ (where $l$ labels the quantum eigenstates of $L^2$), this amounts to the definition of a dimensionless,
small perturbative parameter $\varepsilon=\tau B$. This definition, together
with a rescaling of time $s=t/\tau$ (such that $s\in [0,1]$ during the pulse) leads to an equation suitable for the application of
time-dependent unitary perturbation theory \cite{daems,sugny} 
\begin{equation} \label{eqn2}
i\frac{\partial}{\partial s}\psi(\theta,\phi,s)=[\varepsilon
L^2-E(s)\cos\theta]\psi(\theta,\phi,s) ,
\end{equation}
where $E(s)=\mu_0\tau F(\tau s)$. At lowest order in $\varepsilon$, the emerging 
dynamical picture is the following \cite{henriksen,sugny} : an individual HCP imparts a
kick to the molecule described by an effective instantaneous evolution operator
$e^{iA\cos\theta}$, where $A=\int_0^1E(s)ds$ is the total pulse area. Between
two kicks, the molecule evolves under the effect of its field-free rotation $e^{-i\varepsilon L^2s}$.

The goal of the field driven molecular orientation is to maximize (or to minimize, depending on the choice of the orientation) for the longest
time duration, the expectation value $\langle \cos\theta\rangle (s)=\langle
\psi(\theta,\phi,s)|\cos\theta|\psi(\theta,\phi,s)\rangle$ after the pulse is over. An understanding of this process can be obtained by analysing the molecular dynamics in a finite subspace $\mathcal H _m^{(N)}$ generated by the first
$(N+1)$ eigenstates of $L^2$, i.e. $|l,m\rangle$ ($l=|m|,|m|+1,...N+|m|$) for a molecule initially in the state $|l_0\geq m,m\rangle$. The
justification of such a reduction is in relation with the finite amount of
energy that a finite number of HCPs of a given area $A$ can transfer to the molecule.
The mathematical advantage it offers is the consideration of an operator 
\begin{equation} \label{eqn4}
C_m^{(N)}=P_m^{(N)}\cos\theta P_m^{(N)} ,
\end{equation}
($P_m^{(N)}$ being the projector on the subspace $\mathcal H _m^{(N)}$) which,
as opposed to $\cos\theta$, has a discrete spectrum. It
turns out that the state $|\chi_m^{(N)}\rangle$, which maximises the orientation
in the subspace $\mathcal H _m^{(N)}$, is the eigenstate of $C_m^{(N)}$ with the
highest eigenvalue. Using the approximation $\langle l,m|\cos\theta|l\pm
1,m\rangle \simeq 1/2$ valid for $l\gg m$, straightforward algebra leads to 
\begin{equation} \label{eqn5}
\left|\chi_m^{(N)}\right \rangle\simeq
\Big(\frac{2}{N+2}\Big)^{1/2}\sum_{l=|m|}^{l=|m|+N}\sin\Big(\pi\frac{l+1-|m|}{N+2}\Big) \ |l,m\rangle  .
\end{equation}
Since the quantum number $m$, related with the azimuthal angle is conserved, we will not write it explicitely, unless necessary. The maximal orientation in this subspace is found to be 
\begin{equation} \label{eqn6}
\left \langle\chi_m^{(N)}\left|\cos\theta \right|\chi_m^{(N)}\right \rangle\simeq \cos\Big(\frac{\pi}{N+2}\Big)  .
\end{equation}
\begin{figure}
\includegraphics[scale=0.4]{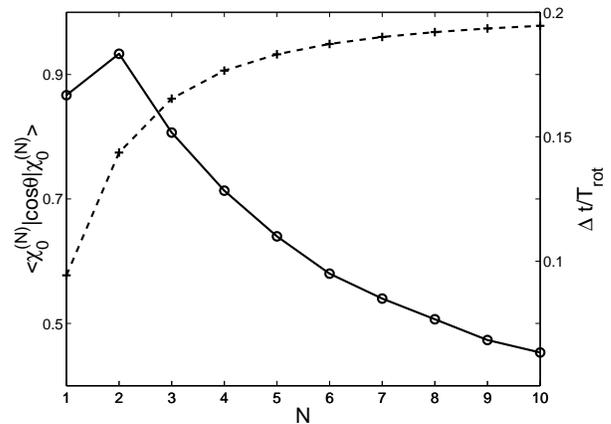}
\caption{\label{figure1} Maximal orientation efficiency (crosses) and associated duration (open circles) as a function of $N$, where $N+1$ is the dimension of the rotationally excited
subspace $\mathcal H_0 ^{(N)}$ (see text). The solid and dashed lines are just to
guide the lecture.}
\end{figure}
For a temperature $T=0 \ \textrm{K}$ ($m=0$), Fig. \ref{figure1} gathers two informations relevant for the
characterization of orientation as a function of $N$; namely
$\langle\chi^{(N)}|\cos\theta|\chi^{(N)}\rangle$ which is the
maximum efficiency [approximatively given by Eq. (\ref{eqn6})] that can ideally be expected for a process that stays confined within the finite subspace $\mathcal H^{(N)}$, and $\Delta t/T_{{\rm rot}}$ which
measures the relative duration of the orientation over which
$\langle\cos\theta\rangle$ remains larger than 0.5 during the field-free
evolution of $|\chi^{(N)}\rangle$. The results, expressed as a fraction of the
rotational period $T_{{\rm rot}}$, are molecule independent. From Fig. \ref{figure1}, we observe that, in order to keep a duration of the order of ${^1/_{10}}$ of the rotational
period (which may amount to durations exceeding 10 ps, for heavy diatomics like e.g. 
NaI), $N$ has to be limited to 5 or 6, which seems rather
limiting. But this turns out to be sufficient for very efficient orientation.
$N=4$ already allows an orientation efficiency larger than 0.91.

\begin{figure}
\includegraphics[scale=0.4]{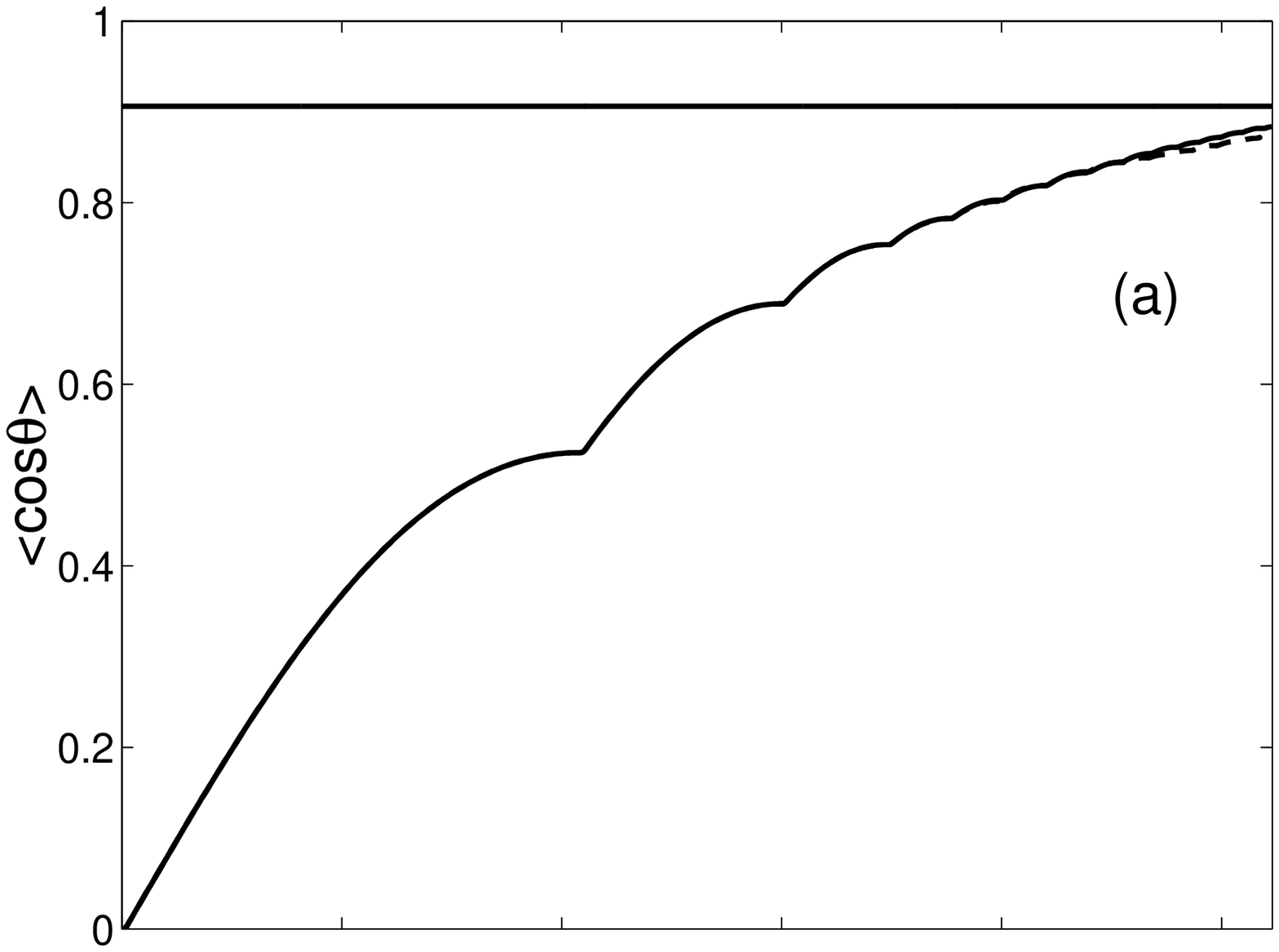}
\includegraphics[scale=0.4]{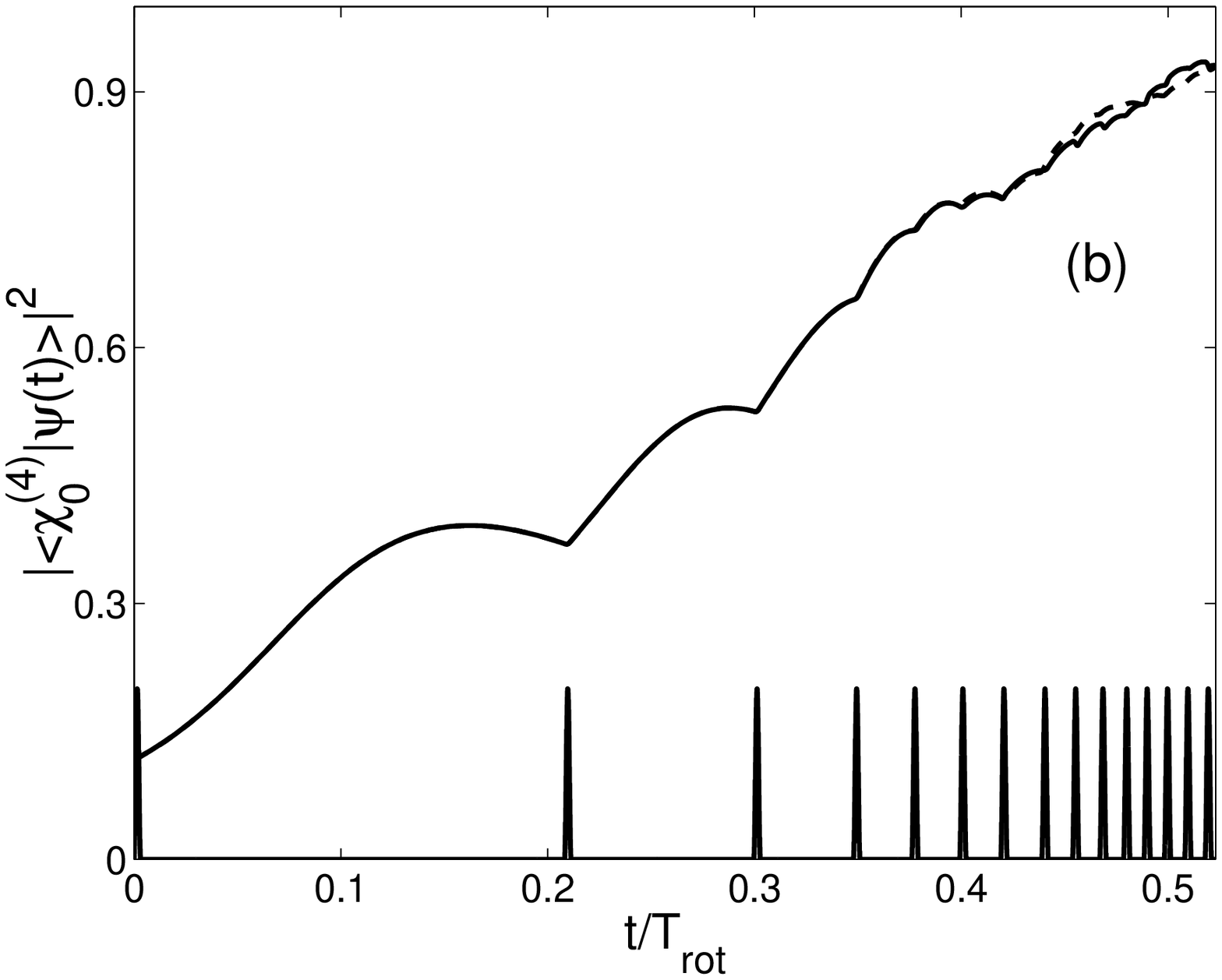}
\caption{\label{figure2} Orientation dynamics during the train of HCPs at $T=0\ \textrm{K}$: panel
(a) for $\langle\psi(s)|\cos\theta|\psi(s)\rangle$ and panel (b) for
$|\langle\chi_0^{(4)}|\psi(s)\rangle|^2$. The solid line corresponds to
$|\psi(s)\rangle$ calculated exactly and the dashed line to the wave function propagated in the subspace $\mathcal H_0^{(4)}$. The train of HCPs is displayed on panel (b) and the optimal orientation is indicated by an horizontal line on panel (a).} 
\end{figure}

Two basic questions are in order : which set ot parameters and which number of kicks have to be chosen to (approximately) remain in the subspace $\mathcal H^{(N)}$, 
and which strategies have to be followed to reach the maximum possible efficiency
within this subspace. The first question is in relation with the kick momentum
transfer operator $e^{iA\cos\theta}$, which is the only evolution operator (as
opposed to the free evolution) that rotationally excites the system and forces
it to expand on a larger subspace. We can estimate the loss outside $\mathcal H^{(N)}$ assuming a preliminar convergence
to $|\chi^{(N)}\rangle$ (as shown by Fig. \ref{figure2}) by looking for the smallness of the norm  
\begin{equation} \label{eqn7}
\left\|\left(e^{iA\cos\theta}-P^{(N)}e^{iA\cos\theta}P^{(N)}\right)\left|\chi^{(N)}\right\rangle\right\|^2=\eta  ,
\end{equation}
which, for small $A$, amounts to 
\begin{equation} \label{eqn8}
 \eta\simeq \frac{\left(A\pi\right)^2}{2(N+2)^3}  .
\end{equation}
This allows us to establish a relation between $A$ and $N$ for a given loss.
$N\simeq 4$ is found compatible with an
$\eta\simeq 0.02$ (not more than 2\% of the rotational population leaving the
subspace $\mathcal H ^{(4)}$) as far as $A$ does not exceed 1.

The second question can be answered by adapting the strategy suggested
for the orientation of a 2D rotor in Ref. \cite{averbukh}, which consists in applying 
laser pulses each time $\langle\cos\theta\rangle$ reaches its maximum. The following argument shows that, if the dynamics stays within the subspace $\mathcal H^{(N)}$, such a strategy
precisely converges to an optimal state $|\chi^{(N)}\rangle$. This is done by
approximating the operators $\cos\theta$ and $e^{iA\cos\theta}$ by $C^{(N)}$ and
$e^{iAC^{(N)}}$ respectively. The interaction with a sudden HCP only alters the
slope of $\langle C^{(N)}\rangle(s)$ and not its value as is clear from the
following relation
\begin{equation} \label{eqn9}
\left \langle e^{-iAC^{(N)}}C^{(N)}e^{iAC^{(N)}}\right \rangle =\left \langle C^{(N)}\right \rangle  .
\end{equation}
Moreover, if a sudden pulse is applied at a time $s_i$ when $\langle
C^{(N)}\rangle (s)$ reaches its maximum $C_i=\langle C^{(N)}\rangle (s_i)$, the
slope undergoes a change from zero to a finite value 
\begin{eqnarray} \label{eqn10}
 \begin{array}{l}
\frac{d}{ds}\langle C^{(N)}\rangle\big|_{\scriptstyle{s_i-0}} =i\left \langle \left[\varepsilon
L^2,C^{(N)}\right]\right \rangle =0, \\
\frac{d}{ds}\langle C^{(N)}\rangle\big|_{\scriptstyle{s_i+0}} =i\left \langle
e^{-iAC^{(N)}}\left[\varepsilon L^2,C^{(N)}\right]e^{iAC^{(N)}}\right \rangle \neq 0 .\\
\end{array}\nonumber\\
\end{eqnarray}
As $\langle C^{(N)}\rangle$ is a periodic, continuously differentiable function, it will reach within the rotational period, a maximum value larger than the one obtained prior to the application of
the pulse. Iterating the strategy, we get an increasing but bounded and
therefore convergent sequence of $C_i$'s. Its limit is a fixed point
$C_i=C_{i+1}$, corresponding to the eigenvectors of the impulsive propagator
$e^{iAC^{(N)}}$ which are also the ones of $C^{(N)}$. Indeed, at a fixed point,
the slopes before and after the interaction with the last pulse have to be 
zero.

Figure \ref{figure2} gives two different views of the orientation
dynamics under the effect of a train of HCPs, separated by time delays corresponding to the above discussed strategy
of maxima, with identical durations
$\varepsilon=0.01$ and pulse areas $A=1$ (that is about 0.3 ps and a field amplitude of $1.5\cdot 10^5$ V cm$^{-1}$ for LiCl \cite{dion2}), which leads to a dynamics that remains within the subspace $\mathcal H^{(4)}$ for the considered numbers of kicks.
From panel (a) it is interesting to note that a single
kick produces an orientation of about 0.5, whereas the appropriate application
of 15 kicks increases this efficiency up to 0.89, which is almost the
optimal limit as found from Fig. \ref{figure1}. On the other hand, the comparison of the average $\langle \cos \theta \rangle$ calculated with the exact wave function
$|\psi(s)\rangle$ and with the one propagated in the subspace $\mathcal H_0^{(4)}$, are
close enough to support the claim that the rotational dynamics actually resides
within $\mathcal H_0^{(4)}$. Panel (b) shows the way the wave function
$|\psi(s)\rangle$ gets close to the optimally oriented state
$|\chi_0^{(4)}\rangle$, showing thus the succesfull outcome of the process. The difference between the two dynamics can be also estimated by Eq. (\ref{eqn8}).
Here again, the close convergence of the two calculations shows a
coherent choice of $A$, $N$ and the number of kicks for appropriately describing the dynamics.
\begin{figure}
\includegraphics[scale=0.4]{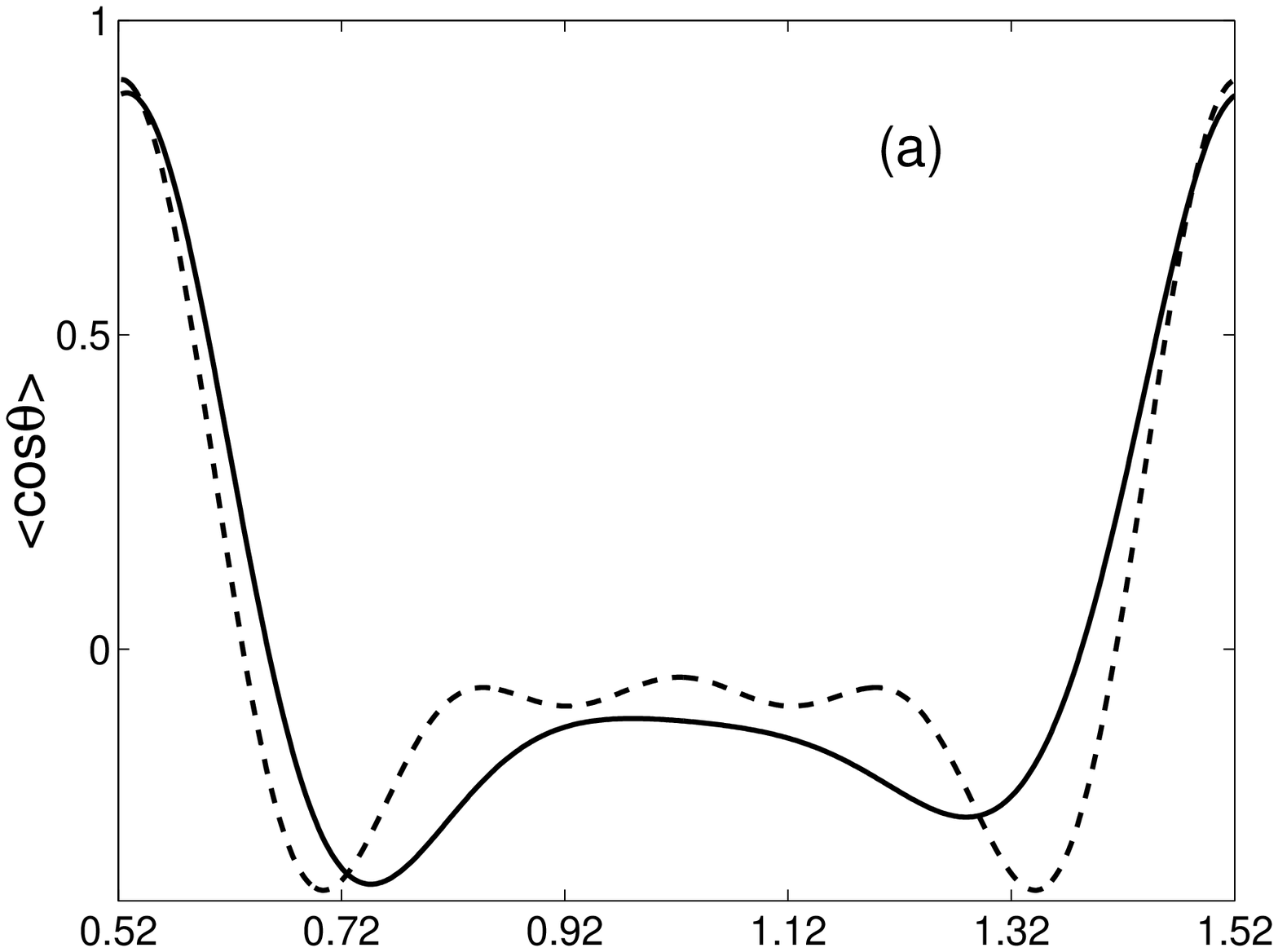}
\includegraphics[scale=0.4]{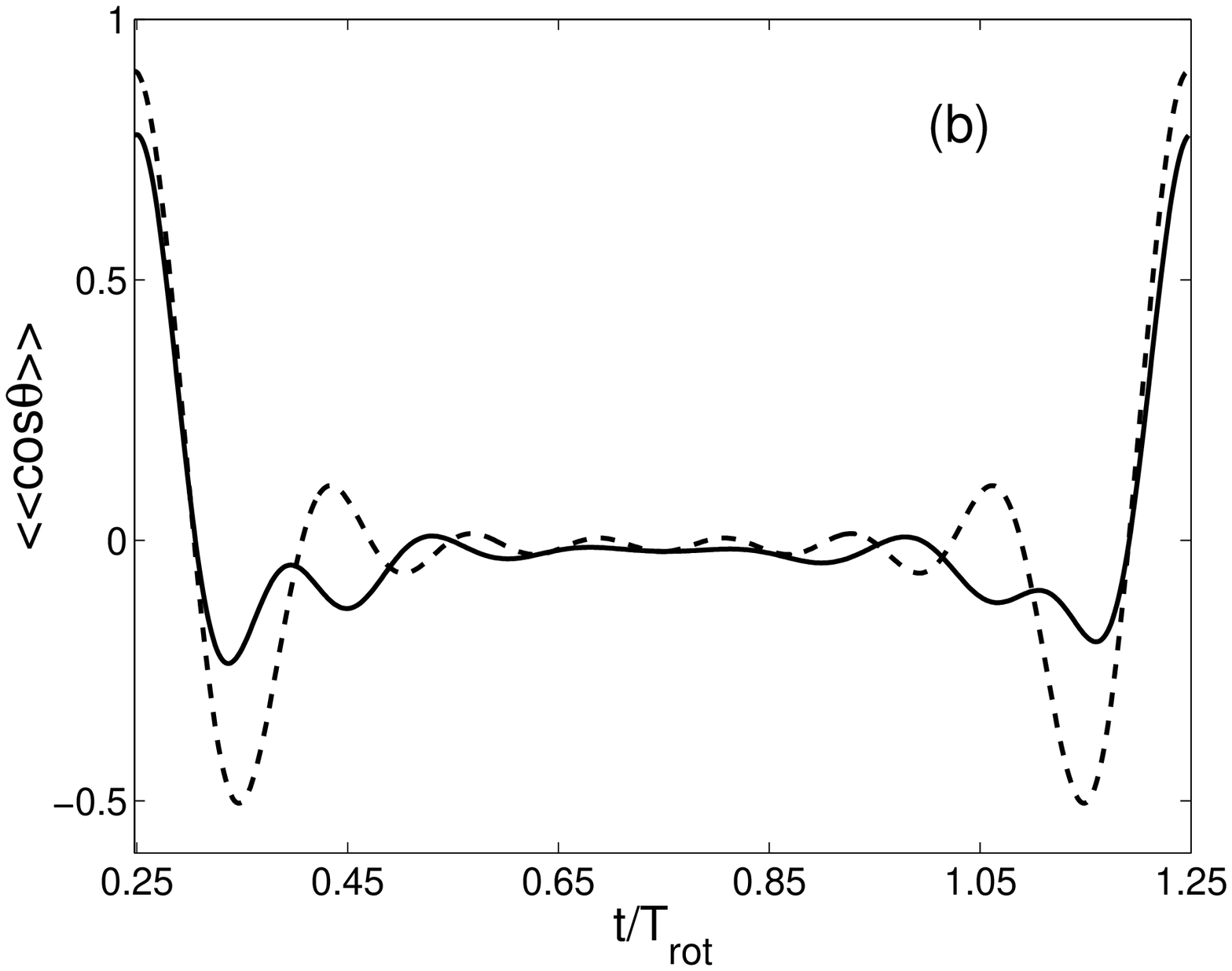}
\caption{\label{figure3} Post-pulse orientation dynamics of the molecule LiCl after interaction with a train of HCPs (the time $t=0$ corresponds to the first kick) in the case $A=1$, $T=0\ \textrm{K}$ [panel (a)] and $A=2$, $T=5\ \textrm{K}$ [panel (b)]. Solid and dashed lines correspond respectively to the averages calculated with the exact wave function and the optimal state [$|\chi_0^{(4)}>$ on panel (a) and $\rho^{(7)}$ on panel (b)].}
\end{figure}
The post-pulse dynamics, which is our main concern, is displayed on Fig.
\ref{figure3} [panel (a)]. A result expected from the previous analysis, but particularly remarkable with respect to previous proposals, is obtained with an efficiency of about 0.89 and a duration of the order of ${^2/_{10}}$ of the rotational period (that
is about 2 ps for a light molecule like LiCl and 20 ps for a heavy one, like
NaI).

Orientation is subject to a drastic decrease with temperature
\cite{dion3,malchholm}. This is basically due to the fact that, with non-zero
temperature, the initial state is a superposition of a statistical ensemble of
rotational states with $m\neq 0$, which tends to misalign the molecule. The efficiency of the 
orientation is characterized by an additional average of
$\langle\cos\theta\rangle$ over the density operators $\rho_m(s)$  
\begin{equation} \label{eqn11}
\langle\langle\cos\theta\rangle\rangle(s)=\sum_{m \in Z} \textrm{Tr}[\rho_m(s)\cos\theta]  .
\end{equation}
We recall that $\rho_m(s)$ evolves according to the von Neumann equation 
\begin{equation} \label{eqn12}
\frac{d}{ds}\rho_m(s)=i[\rho_m(s),\varepsilon L^2-E(s)\cos\theta],
\end{equation}
with as initial condition 
\begin{equation} \label{eqn13}
\rho_m(0)=\frac{1}{\mathcal Z}\sum_{l\geq |m|}\left|l,m\right \rangle e^{-Bl(l+1)/k T}\langle l,m|,
\end{equation}
where $\mathcal Z=\sum_{m\in Z}\sum_{l\geq|m|}e^{-Bl(l+1)/kT}$ is the partition
function and $k$ the Boltzmann constant. Following our previous analysis, we
are looking for the optimal density operator which maximizes $\langle\langle\cos\theta\rangle\rangle$
in the subspace $\mathcal H^{(N)}$, which is given by 
$\rho^{(N)}=\sum_m\rho_m^{(N)}$ where $\rho_m^{(N)}=|\chi_m^{(N)}\rangle
\textrm{Tr}[\rho_m^{(N)}]\langle \chi_m^{(N)}|$ with the constraint related to the conservation
of $m$, expressed as $\textrm{Tr}[\rho_m^{(N)}]$ to be kept constant 
\begin{equation} \label{eqn14}
\textrm{Tr}[\rho_m^{(N)}]=\frac{1}{\mathcal Z}\sum_{|m|}^{l=|m|+N}e^{-Bl(l+1)/kT}  .
\end{equation}
The effect of a sudden pulse on $\rho$ is $e^{-iA\cos\theta}\rho
e^{iA\cos\theta}$, and because its optimal value corresponds to a fixed point of
the $\langle\langle C^{(N)}\rangle\rangle(s_i)$ sequence, it is precisely the application of a train
of HCPs with individual pulses at times $s_i$ where $\langle\langle \cos\theta \rangle\rangle (s_i)$
reaches its maximum, that converges to the best possible orientation within
this model. The resulting dynamics is plotted in Fig. \ref{figure3} [panel (b)], with a
maximum efficiency of about 0.75 and a duration of about ${^1/_{20}}$ of the
rotational period. To our knowledge, this is the largest duration and efficiency achieved up to date for a thermal ensemble.

In conclusion, we have presented tools for controlling 
molecular orientation dynamics using a train of HCPs, achieving both efficiency and duration of the post-pulse orientation. Moreover, this scheme can be expected to be transposable to a generic system, with free periodic
dynamics governed by an Hamiltonian $H_0$, and for which we are aiming to
optimally control an observable $\mathcal O$ (i.e. maximize or minimize the
average $\langle\mathcal  O\rangle(t)$ of an upper or lower bounded operator
$\mathcal O$ which does not commute with $H_0$). This could be done through a
device that perturbs the system according to a unitary operator $U$, which
commutes with  $\mathcal O$, such that its application does not alter  $\langle
\mathcal O\rangle=\langle U^{-1}\mathcal O U\rangle$ on one hand, and the
optimal target state is an eigenfunction of both $\mathcal O$ and $U$ on the
other hand. This optimum corresponds to a fixed point of the sequence
$O_i=\langle\mathcal O\rangle(t_i)$ where $t_i$ are the times when
$\langle\mathcal O\rangle(t)$ reaches its maximum (or minimum) under the free
evolution. In particular, this scheme provides a comprehension of previous
works on alignment and orientation of 2D and 3D rotors \cite{Leibscher,averbukh}.

\end{document}